\documentclass[
 amsmath,amssymb,
pra,twocolumn
]{revtex4-2}
\usepackage{graphicx}
\usepackage{dcolumn}
\usepackage{bm}

\usepackage{xcolor}

\begin{document}


\title{\textbf{Dual-rate optical frequency comb based on a cylindrical microresonator with a nanoscale radius variation} }%



\author{Alena Kolesnikova$^{1,2}$}
\email[Corresponding author: ]{a.kolesnikova@skoltech.ru}
\author{Andrey Gelash$^{2}$}
\author{Ilya Vatnik$^{1}$}

\affiliation{$^{1}$Novosibirsk State University, Novosibirsk, Russia\\$^{2}$Skolkovo Institute of Science and Technology, Moscow, Russia}

\date{\today}

\begin{abstract}
We theoretically find that a cylindrical microresonator with nanoscale radius variation can be a source of coherent microcombs with two repetition rates, independently controlled by the device geometry. This phenomenon arises from the formation of two-dimensional axial-azimuthal dissipative Kerr solitons (2D DKS), which we discovered through numerical simulations of the nonlinearly coupled Lugiato-Lefever equations with realistic microresonator device parameters. We demonstrate that, despite the variation in the dispersion characteristics of azimuthal modes at different axial numbers, the spontaneous synchronization mechanism leads to a unified soliton repetition rate. We consider near-parabolic variations in the radius of a cylindrical resonator, compatible with the surface nanoscale axial photonics (SNAP) platform. In our example, we generate 2D DKS with an azimuthal rate of approximately 329 GHz and an axial rate of 6 GHz, which can be adjusted on demand by selecting the appropriate resonator geometry, thereby promising compact sources of frequency combs operating simultaneously across different frequency scales.
\end{abstract}

\maketitle

{\it Introduction.} The discovery of dissipative Kerr solitons (DKS) in optical microresonators represents a paradigm shift in nonlinear optical science and the fabrication of miniature and microchip-integrated photonic devices \cite{vahala2003optical,Strekalov_2016,Kipp11,Herr2012,herr2014temporal}. These coherent nonlinear structures emerge from the balance between dispersion, Kerr nonlinearity, loss, and external driving in high‑quality‑factor (Q) resonators. Their finding has enabled the generation of coherent, broadband optical frequency microcombs, offering compact and energy‑efficient alternatives to traditional mode‑locked laser systems \cite{Herr2012,herr2014temporal,lucas2018spatial,KippScience2018,gaeta2019photonic,Fujii:22,fujii2023versatile}.

Bulk crystalline and on-chip fabricated microresonators are the two main platforms for reliable generation of DKS, which repetitive motion creates coherent and equidistantly spaced frequency microcombs. While on-chip microcomb devices promise scalability and integration with other components of various optical systems \cite{herr2026frequency}, the crystalline microresonators are easy to fabricate and exhibit exceptionally high Q that exceeds $10^{10}$ \cite{kippenberg2011microresonator}. They also provide a robust testbed for exploring new phenomena in soliton physics, as demonstrated by a series of fundamental studies conducted on this platform, including recent developments, which altogether laid the groundwork for understanding dissipative soliton regimes \cite{lucas2017breathing,pavlov2017soliton,guo2017intermode,wang2025collision}.

Bulk microresonators, often based on doped crystals or optical fibers, can have different geometries, such as spherical, toroidal, or disk-shaped. They are typically from micrometer to millimeter scale and made of materials such as magnesium fluoride (MgF\(_2\)) or crystalline quartz. The extreme optical confinement and enhanced Kerr nonlinearity in these devices allow soliton formation at milliwatt‑level pump powers. Remarkably, different types of microresonators are governed by similar underlying theoretical frameworks, most notably the Lugiato–Lefever equation (LLE) \cite{lugiato1987spatial,kaup1978solitons,nozaki1986low,wabnitz1996control}. The equation describes the nonlinear dynamics of the azimuthal mode field, whose resonant frequencies are nearly equidistant, forming solitons that run along the perimeter of the microresonator. The repetition rate of the resulting comb at the system's output is determined by the round-trip time of such a soliton.

The Surface Nanoscale Axial Photonics (SNAP) platform enables the reliable fabrication of optical fiber-based microresonators with ultrafine nanoscale variations in the effective radius spanning hundreds of microns along the fiber axis. Proposed over a decade ago \cite{Sumetsky:12}, this technology is continually advancing and now offers high-precision sub-angstrom optical control and reliable identification methods \cite{savchenkov2026prism}. Theoretical studies predict that SNAP microresonators can support the generation of conventional azimuthal DKS with a stationary axial distribution \cite{Kartashov:18} and axial DKS with a stationary distribution across the cylinder cross section \cite{Kolesnikova:25}. Axial solitons propagate along the cylinder axis and are repeatedly reflected from the walls of the effective potential created by the radius variation. Due to the low average propagation velocity of such solitons, the repetition rate of the axial comb spectrum is hundreds of MHz, which is several orders of magnitude lower than in the azimuthal case. Recent studies have demonstrated the reliable generation of azimuthal Kerr combs in cylindrical SNAP microresonators \cite{Eadie:25}, promising the realization of fully coherent DKS states in the near future.

In this work, we propose a realistic SNAP microresonator configuration that supports both azimuthal and axial modes, thanks to the spontaneous formation of 2D DKS that propagate in a thin layer of bulk material beneath the cylinder surface (see Fig. 1).  For theoretical study, we suggest a system of coupled Lugiato-Lefever equations, which describes the nonlinear interaction between axial-azimuthal modes. We design a localized effective potential using nearly parabolic nanoscale variations in the radius of the cylinder to create three axial modes coupled nonlinearly with each other. 

As a result of modeling such a system, we observe the spontaneous formation of two-dimensional axial-azimuthal dissipative Kerr solitons, producing a dual-rate optical frequency comb. While the azimuthal soliton dynamics generate high repetition rates of approximately 329 GHz, the azimuthal soliton motions are much slower and produce a detectable repetition rate of 6 GHz. This dual set of optical lines can be adjusted on demand by choosing the appropriate resonator geometry, promising compact sources of frequency combs operating simultaneously at different frequency scales.



\begin{figure}
    \centering
    \includegraphics[width=\linewidth]{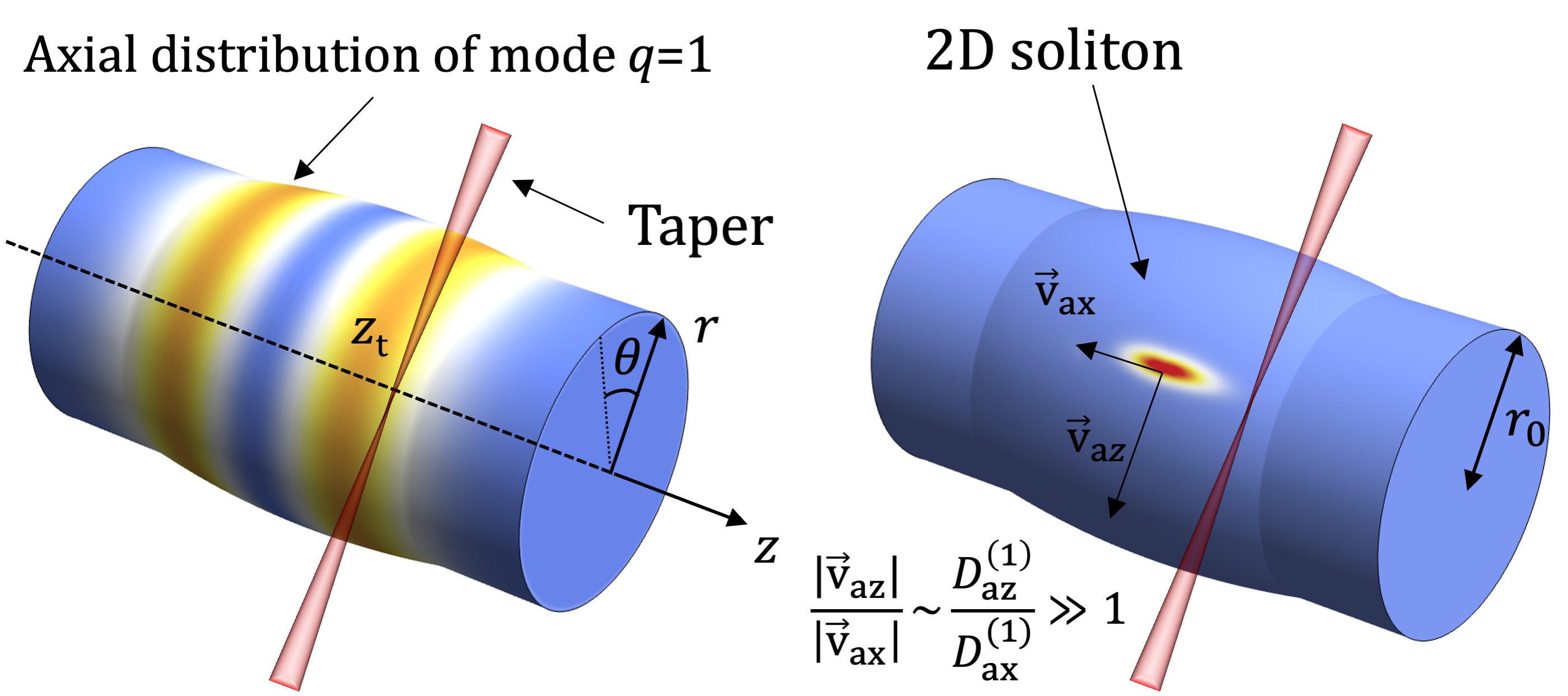}
    \caption{Schematic representation of the SNAP microresonator with the input/output taper positioned in the point \(z_t\). The resonator surface is coloured according to the intensity distribution of the mode with the axial number \(q=1\) and the two-dimensional axial-azimuthal dissipative Kerr soliton.}
    \label{fig:snap}
\end{figure}

{\it Model of cylindrical SNAP microresonator}.
The electric field $E(\vec r, t)$ with linear TE or TM polarization inside a SNAP microresonator can be decomposed into axial-azimuthal modes \cite{Kolesnikova:24}:
\begin{equation}
    E(\vec r, t) = \sum_{q,m}a_{q,m}(t)e_{q,m}(\vec r)e^{-i\omega_q^m t}.
\end{equation}

Here \(a_{q,m}(t)\) is the amplitude of the modes with axial quantum number \(q\) and azimuthal quantum number \(m\) and resonance frequency \(\omega_q^m\). The spatial distribution of the axial-azimuthal modes is determined by the function \(e_{m,q}(\vec r)=Z_{m,q}(z)e_{m}(r)e^{im\theta}\). The spatial distribution of modes in the cross-section \(e_m(r)\) is described by the Bessel function \cite{Demchenko:13}. For large azimuthal numbers and the same radial number and polarization (we consider radial number \(p=1\) and TE-mode), which are typical for modes near a wavelength of \(1.55\) \textmu m, the spatial distribution is localized near the fiber surface and weakly depends on the number \(m\).

In the first approximation, the resonant frequencies of modes with the same axial number are almost equidistant \(\omega_m^q\approx\cfrac{cm}{n_mr_0}\), therefore we can expand in a series the resonant frequency near the pump mode with \(m_0\): \(\omega_q^m=\omega_q^{m_0}+D_q^{(1)}\mu+D_q^{(2)}\mu^2\), where \(D_q^{(1)}\) is the azimuthal free spectral range, \(D_q^{(2)}\) is the second order azimuthal mode dispersion, \(\mu=m-m_0\) is the ordinal number from the central azimuthal mode resonances.


The spatial distribution of modes along the \(z\)-axis is affected by the small size of radius variation \(\Delta r(z)\ll r_0\) and can be described using the stationary Schrödinger equation \cite{Sumetsky:11,Kolesnikova:23}:
\begin{equation}
    \cfrac{d^2 Z(z)}{d^2z}+2k_{m}^2\cfrac{\Delta r(z)}{r_0}Z(z)=2k_{m}^2K_m\cfrac{\omega-\omega_{m}}{\omega_{m}}Z(z), \label{Z_eq}
\end{equation}
where \(k_{m}=\cfrac{\omega_{m}n_{m}}{c}\), \(K_m = 1+\cfrac{\omega_{m}}{n_{m}}\cfrac{\text{d}n}{\text{d}\omega}(\omega_{m})\). The solution to the equation is a set of eigenvectors - stationary distributions of modes along the \(z\) axis of the cylinder \(Z_{m,q}(z)\) and the corresponding resonant frequencies \(\omega_q^m\).

In turn, by choosing a near-parabolic variation of the radius \(\Delta r(z) = \delta r\left(1-\left(2z/z_0\right)^2+\alpha\left(2z/z_0\right)^4\right) \), by analogy with the known solution for a harmonic oscillator, we can obtain an almost equidistant spectrum of resonances corresponding to one azimuthal number and different axial numbers. Here \(\delta r\) is the maximum value of radius variation, \(z_0\) is the half-width of radius variation, \(\alpha\) is the anharmonicity coefficient. For \(\alpha=0\) exact solution of Eq.\eqref{Z_eq} is \(\omega_q^m=\omega_m-\omega_m\cfrac{\delta r}{r_0}+\cfrac{c}{K_m n_m}\left(\cfrac{2\delta r}{z_0^2}\right)^{1/2}(q+1/2)\). 

We can also expand the resonant frequencies by axial number near the mode with \(q_0\) and \(m_0\) into which pumping will be carried out: \(\omega_q^{m_0}=\omega_{q_0}^{m_0}+\Delta_{\text{ax}}l+D_{\text{ax}}^{(2)}l^2\), where \(\Delta_{\text{ax}}\) is the axial free spectral range, \(D_{\text{ax}}^{(2)}\) is the second order axial mode dispersion, \(l=q-q_0\) is the ordinal number from the pumped axial mode resonance.



By analogy with quantum mechanics \cite{landau1958quantum}, in a potential well of finite depth, the complete basis of the space of functions describing the axial field distribution \(Z_{m,q}(z)\) is a finite discrete set of wave functions corresponding to bound states—axial modes—and modes of the continuous spectrum \(e^{i\beta z}\). To describe the axial distribution localized in a microresonator, we neglect the losses due to radiation escaping into the continuous spectrum, scattering at infinity, and consider only the discrete set of localized axial modes.

To describe the 2D nonlinear dynamics, we derived nonlinearly coupled equations for the amplitudes of axial modes \(A_q(\theta,t)\) under the assumption that the spatial distribution of the field does not depend on the number of the azimuthal mode for large \(m\) \(e_{m}(r)\equiv e( r)\), \(Z_{m,q}(z)\equiv Z_{q}(z)\). Thus, the electric field can be represented as:

\begin{eqnarray}
\label{E_via_Aq}
    &&E(\vec r,t) = \\\nonumber
    &&\sqrt{\cfrac{2\hbar\omega_L}{n_L^2\varepsilon_0}}\sum_{q}A_q(\theta,t)e( r)Z_q(z)e^{-i\omega_q^{m_0}t+i\omega_Lt}+\text{c.c.}
\end{eqnarray}
Here \(A_q(\theta,t)=\sqrt{\cfrac{n_L^2\varepsilon_0}{2\hbar\omega_L}}\sum_m a_{q,m}(t)e^{im\theta}\) - slowly varying amplitude of the axial mode \(q\), \(|A_q|^2\) is the photon number. The mode profile normalized as \(\int|e(r)Z_q(z)|^2d^3\vec r=1\), \(\omega_L\) is the pump frequency, \(n_L=n(\omega_L)\) is the refractive index. 

In this assumption, we obtain the system of coupled nonlinear equations for an arbitrary number of azimuthal and axial modes of the SNAP microresonator:

\begin{eqnarray}
\label{General}
    \cfrac{\partial A_q}{\partial t}=\left[-\cfrac{\kappa_q}{2}-i\Delta_q+\Delta D^{(1)}_{q}\cfrac{\partial}{\partial\theta}+iD_q^{(2)}\cfrac{\partial^2}{\partial\theta^2}\right]A_q \quad\quad
    \\\nonumber +i\sum_{i,j,k} \delta(i-j+k-q)g_{ijkq}A_j^*A_iA_k +\delta_{0q_0}\sqrt{\kappa^{(c)}_q}A_{\text{in}}
\end{eqnarray}
Here \(\kappa_q=\Gamma+\kappa^{(c)}_q\) is the total mode losses, consisting of the microresonator's own losses \(\Gamma\) and the resonance coupling losses with the taper \(\kappa^{(c)}_q\), which  depend on the taper position \(z_t\): \(\kappa^{(c)}_q=C|Z_q(z_t)|^2\), \(C\) is a coupling parameter set by the overlap of taper and resonator fields \cite{Kolesnikova:23}. To derive Eq.~(\ref{General}) we switched to the reference frame of the central axial mode \(q_0\), so that \(\Delta D^{(1)}_{q}=D_{q}^{(1)}-D_{q_0}^{(1)}\), and subtracted the rapidly oscillating frequency associated with the free spectral range of axial modes \(A_q(\theta,t)\equiv \mathcal{F}^{-1}\left(A_q(\mu,t)e^{iD_{q_0}^{(1)}\mu t}\right)e^{ -i\Delta_{\text{ax}}lt}\), where $\mathcal{F}^{-1}$ is the inverse Fourier transform. The other parameters in the model~(\ref{General}) are
\(\Delta_q=\omega_{q}^{m_0}-\omega_L-\Delta_{\text{ax}}l\) - detuning of the pump frequency \(\omega_L\) from the resonance frequency \(\omega_q^{m_0}\), \(\Delta D^{(1)}_{q}=D_{q}^{(1)}-D_{q_0}^{(1)}\). \(g_{ijkq} = \cfrac{\hbar \omega_L^2n_2}{n_L^2V_{ijkq}}\) is the Kerr nonlinearity coefficient, \(n_2\) is the nonlinear refractive index, \(V_{ijkq} =\int e_i(\vec r)e_j(\vec r)e_k(\vec r)e_q(\vec r) d^3\vec r\). \(A_{\text{in}}=\sqrt{\cfrac{P_{\text{in}}}{\hbar\omega_L}}\), where \(P_{\text{in}}\) is the pump power. Compared to the previously considered 2D LLE model \cite{Kartashov:18}, Eq.~(\ref{General}) is better suited for the description of the two-dimensional nonlinear dynamics in SNAP microresonators by taking into account the dependence of the azimuthal free spectrum range, second-order dispersion and taper losses on the axial number.

{\it SNAP microresonator with three axial modes.} As we demonstrate below, to provide the effective nonlinear interaction between modes with different axial numbers and generate 2D DKS, it is sufficient to consider the case of a radius variation, which leads to the emergence of only three sets of azimuthal modes with different axial distributions. So, a cylindrical microresonator based on a fused silica optical fiber with a radius of \(r_0=100\,\mu\)m was chosen for the simulation to ensure positive azimuthal mode dispersion \(D_q^{(2)}>0\) \cite{Kolesnikova:24,Eadie:25}. The near-parabolic radius variation has the realistic width \(z_0=160\,\mu\)m and the height \(\delta r=10\) nm, which, according to numerical solution of the eigenvalue problem for Eq.~\eqref{Z_eq}, correspond to the three-level potential, that is, a SNAP microresonator with three axial modes. The anharmonicity coefficient \(\alpha=0.25\) provides positive axial mode dispersion \(D_{\text{ax}}^{(2)}/(2\pi)=8.8\,\)MHz, which is necessary for the modulation instability of axial modes \cite{Kolesnikova:24}. This radius variation contains three axial modes with axial number \(q=0,1,2\) (corresponding \(l=1,0,-1\), \(\omega_0^m>\omega_2^m\)) and the free spectral range \(\Delta_{\text{ax}}/(2\pi)=6\,\) GHz. From here on, the numbers \(l\) will be used to designate the axial mode, the pumping axial mode has \(l=0\).



Solving the eigenvalue problem of Eq.~\eqref{Z_eq}, we find the spatial distribution of modes \(Z_l(z)\), with the help of which the effective volumes of modes \(V_{ijkl}\) and the coupling coefficients of the taper with axial modes \(\kappa_l^{(c)}\) were determined, and also resonant frequencies \(\omega_l^m\) from the values of which the free spectral range \(D_l^{(1)}\) and the second-order dispersion of azimuthal modes \(D_l^{(2)}\) were calculated (see Table \ref{table:1} for detailed values).

The resonator intrinsic losses \(\Gamma=10^7 \text{s}^{-1}\) and the coupling strength are \(C= 2\cdot 10^4\,\)m/s chosen to match the experimentally realized parameters \cite{Kolesnikova:23}. The taper position along the resonator axis \(z_t=-81\,\mu\)m corresponds to the condition of the minimum threshold power for the pumping axial mode \(\kappa_0^{(c)}=\Gamma/2\).

\renewcommand{\arraystretch}{1.0}
\begin{table}[h!t]
\begin{tabular}{ |c|c|c|c| } 
 \hline
 ~~~ & ~~~~~$l=1$~~~~~ & ~~~~~$l=0$~~~~~ & ~~~~~$l=-1$~~~~~~ \\
 \hline
 $D_{l}^{(1)}/2\pi, \,\text{GHz}$  & 329.1103 & 329.1077 & 329.1076 \\
 \hline
 $D_{l}^{(2)}/2\pi, \,\text{kHz}$  & 239 & 254 & 256 \\
 \hline
 $\kappa_{l}/2\pi, \,\text{\textmu s}^{-1}$ & 0.33 & 5 &37.6 \\
 \hline
  $g_{l}, \,\text{s}^{-1}$ & 0.0118 & 0.009 &0.007 \\
 \hline
\end{tabular}
\caption{Parameters for modeling system \eqref{eq:three} for different axial numbers.}
\label{table:1}
\end{table}

In the case of three axial modes, the general system \eqref{General} rewrites as three LLE equations of anomalous dispersion coupled via nonlinear terms,
\begin{widetext}
\begin{eqnarray}
\label{eq:three}
    \cfrac{\partial A_0}{\partial t}=\left[-\cfrac{\kappa_0}{2}-i\Delta_0+iD_0^{(2)}\cfrac{\partial^2}{\partial\theta^2}+ig_0|A_0|^2+i2g_{11}|A_1|^2+i2g_{-1-1}|A_{-1}|^2\right]A_0 +2ig_{-11}A_{-1}A_1A_0^{*}+\sqrt{\kappa^{(c)}_0}A_{\text{in}} \quad\quad\quad\quad\\\nonumber  
    \cfrac{\partial A_{\pm1}}{\partial t}=\left[-\cfrac{\kappa_{\pm1}}{2}-i\Delta_{\pm1}+\Delta D^{(1)}_{\pm 1}\cfrac{\partial}{\partial\theta}+iD_{\pm 1}^{(2)}\cfrac{\partial^2}{\partial\theta^2}+ig_{\pm1}|A_{\pm1}|^2+i2g_{-1-111}|A_{\mp1}|^2+i2g_{\pm1\pm1}|A_0|^2\right] A_{\pm 1} + ig_{-11}A_{\mp 1}^{*}A_0^2
\end{eqnarray}

\end{widetext}

%
%
%
    
Here \(g_i\equiv g_{iiii}\), \(g_{\pm i\pm i} \equiv g_{00\pm i\pm i}\), and \(g_{-ii}\equiv g_{00-ii}\).




\begin{figure*}[t!]
    \centering
    \includegraphics[width=\linewidth]{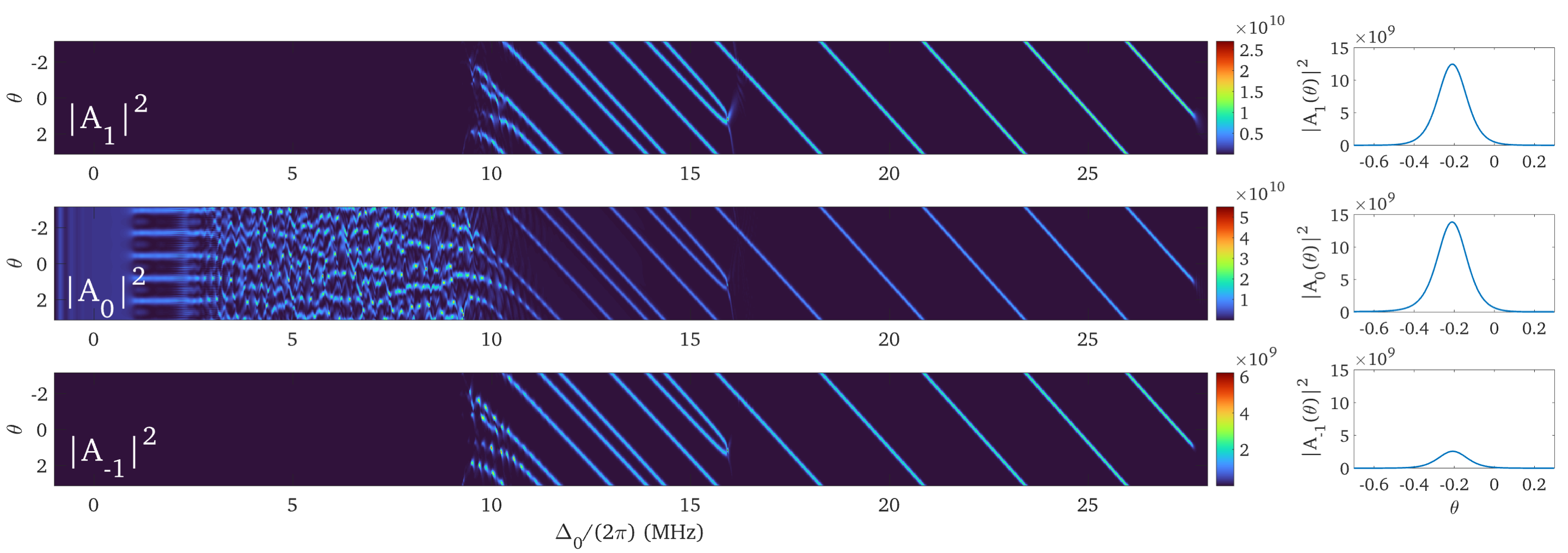}
    \caption{Evolution of the spatial distribution of the field \(|A_l(\theta,\Delta_0(t))|^2\), leading to generation of azimuthal coherent states of the 2D DKS in SNAP microresonators with the three axial modes. Insets show the azimuthal intensity distribution of solitons of each axial number at the pump detuning \(\Delta_0/(2\pi)=27\) MHz.}
    \label{fig:azimuthal_soliton}
\end{figure*}

{\it 2D axial-azimuthal solitons.} We use a standard approach to reach soliton states by pumping energy into the resonator above the modulation instability threshold and slowly tuning the pump frequency into the soliton existence range \cite{herr2026frequency}. We set the pump power \(P_{\text{in}}=0.04\,\)W into the central mode and scan frequency in the region \(\Delta_0/(2\pi)\in[-1\, \text{MHz},\,28\, \text{MHz}]\). We observe the conventional nonlinear dynamics of the LLE model: first, an increase of the interresonator field amplitude; then, the emergence of Turing rolls; and finally, the modulation instability of azimuthal modes with a single axial number, see Fig.\ref{fig:azimuthal_soliton}. At \(\Delta_0/(2\pi)\approx 9.5\,\) MHz, the axial modulation instability triggers effective energy transfer from the central mode with $l=0$ to the side modes with $l=\pm 1$, resulting in the formation of azimuthal solitons in each axial component at further detuning values, see Fig.\ref{fig:azimuthal_soliton}.



Remarkably, velocities of azimuthal solitons from different axial modes synchronize despite the initial difference in the free spectral ranges of the azimuthal modes \(D_l^{(1)}\), so that the resulting repetition rate \(\Delta\nu_{\text{az}}\sim329.1085\) GHz corresponds to the average over the azimuthal modes \(D^{(1)}_{\text{av}}=\sum_l D^{(1)}_l/3\). Recall that equations \eqref{General} and \eqref{eq:three} are written in the reference frame of the central axial mode, the free spectral range of which differs from the repetition frequency of the 2D DKS \(D_0^{(1)}\neq D^{(1)}_{\text{av}}\). Therefore, the azimuthal coherent states of the 2D DKS shown in Figure \ref{fig:azimuthal_soliton} demonstrate drift, corresponding to the repetition frequency \(D^{(1)}_{\text{av}}-D_0^{(1)}\).

\begin{figure}[hptp!]
    \centering
    \includegraphics[width=\linewidth]{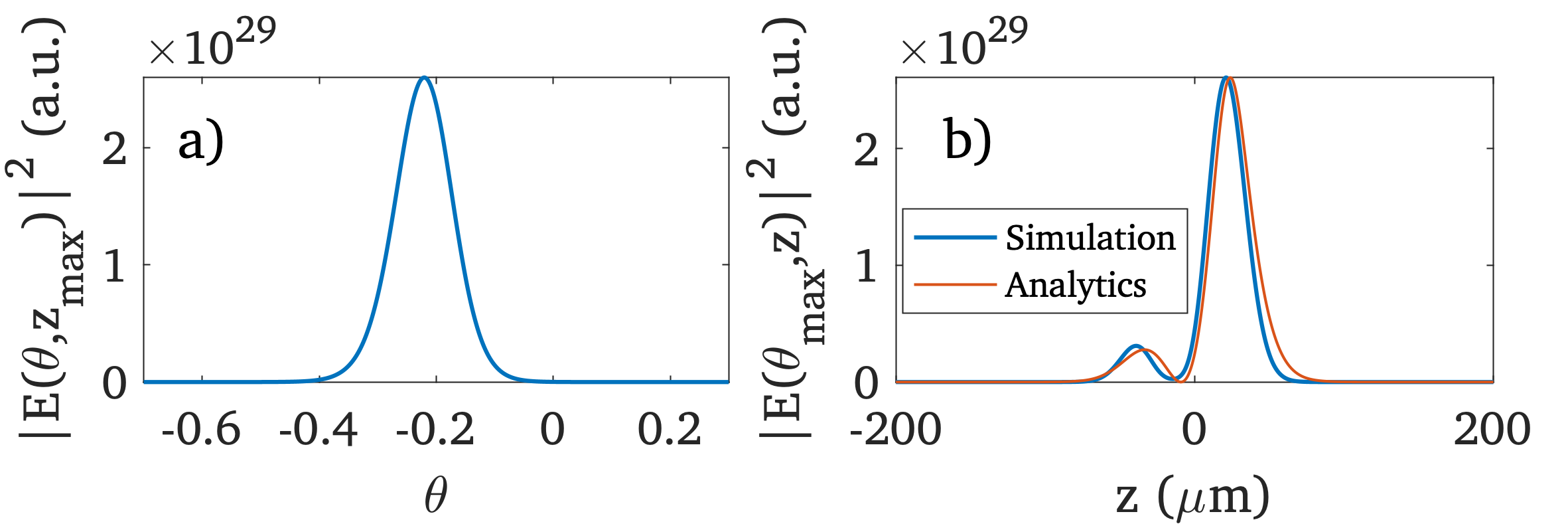}
    \caption{Cross-sections of the 2D soliton intensity distribution. a) Azimuthal \(|E(\theta,z_{\text{max}})|^2\) and b) axial \(|E(\theta_{\text{max}},z)|^2\) intensity distributions at the pump detuning \(\Delta_0/(2\pi)=27\) MHz. Comparison with analytical 1D axial soliton solution obtained in \cite{Kolesnikova:25}, presented for parameters \(A_s = 3.35\cdot 10^{14}\), \(A_0 = 0.56 A_s\), \(z_s = 2.1 \,\mu\)m, \(w_s = 15.5 \,\mu\)m, \(\varphi_s = -0.02\).}
    \label{fig:solitons}
\end{figure}

This synchronization enables the formation of a fully coherent 2D axial-azimuthal soliton localized both in $\theta$ and $z$. To reconstruct the 2D intensity distribution of the soliton, we stop detuning at \(\Delta_0/(2\pi)=27\) MHz and use Eq.~(\ref{E_via_Aq}), substituting the computed $A_l(\theta)$ and integrating over $r$. In Fig.~\ref{fig:snap}, we show a snapshot of the soliton intensity distribution over the resonator, and in Fig.~\ref{fig:solitons}, the intensity cross-sections taken at the point of soliton maximum. We confirm, via additional numerical simulations at fixed detuning, that the 2D soliton is stable, i.e., fully coherent.

The generated 2D DKS exhibits exponential localization in both azimuthal and axial directions, as illustrated by Fig. \ref{fig:solitons}. Thanks to the spatial localization of the axial modes, even three of them are sufficient to reach the characteristic $\cosh^{-1}$ soliton profile in the $z$-direction, which we confirm by comparison of the numerically obtained solution with the approximate analytical 1D axial solution \(|E(\theta_{\max},z)|^2=\left|A_s/\cosh((z-z_s)/w_s)+e^{i\varphi_s}A_0 Z_0(z)\right|^2\), derived in \cite{Kolesnikova:25}. As shown in \cite{Kolesnikova:25}, the shape of the axial soliton is always determined by the predominance of the mode with \(q=1\) featuring two distinct maxima along the cylinder axis, even for the microresonator configurations with a large number of axial modes. Hence, the 2D DKS obtained here represent a general form of localized coherent structure in SNAP microresonators.

\begin{figure}[hptp!]
    \centering
    \includegraphics[width=\linewidth]{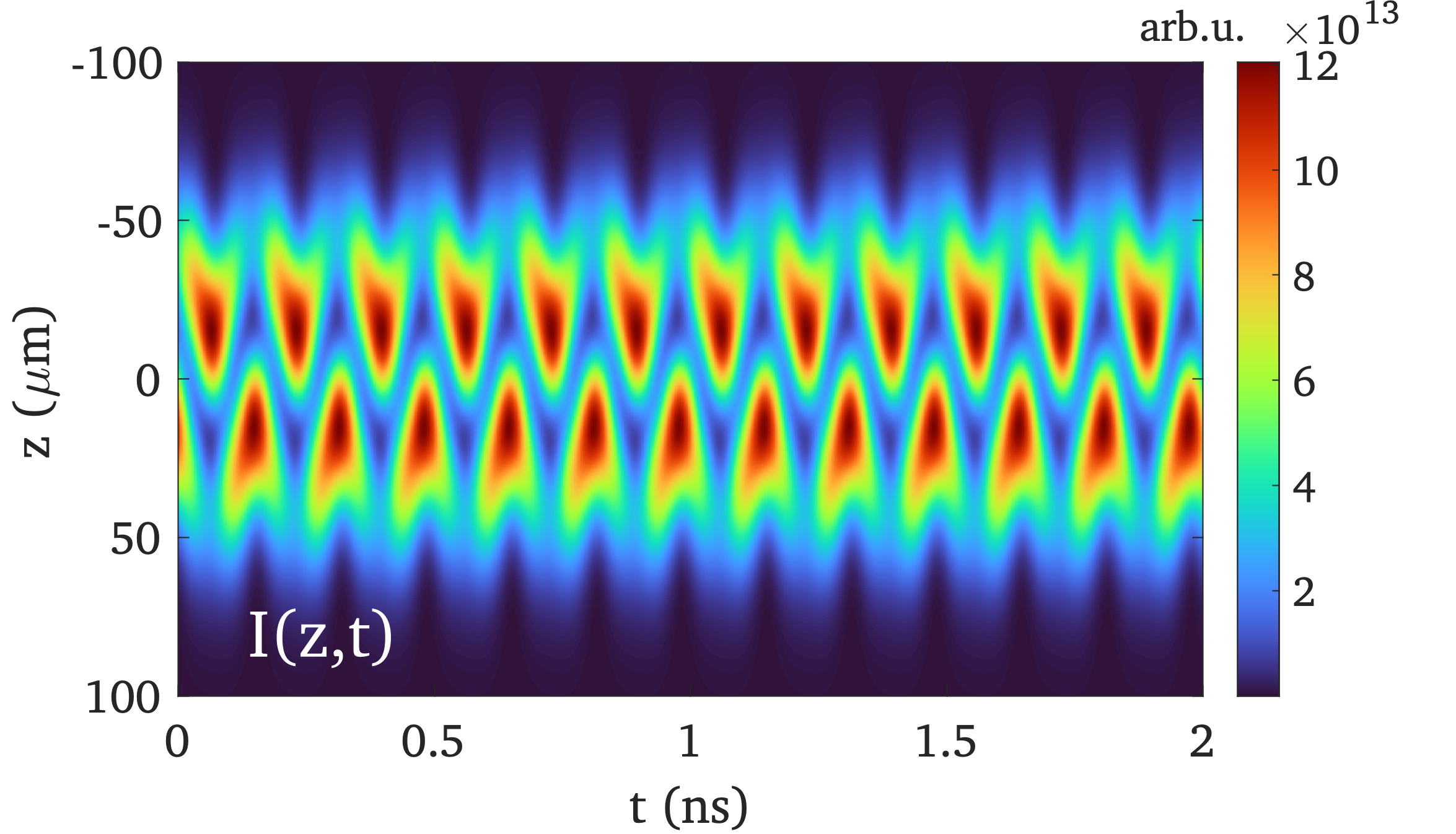}
    \caption{Axial intensity distribution evolution \(I(z,t)\) of the 2D soliton at the pump detuning \(\Delta_0/(2\pi)=27\) MHz.}
    \label{fig:axial_solitons}
\end{figure}

To investigate the axial dynamics of the 2D soliton, we compute the intensity distribution along the resonator axis integrated over the azimuthal coordinate \(I(z,t)=\int{\left|\sum_lA_l(\theta,t)Z_l(z)e^{-i\omega_l^{m_0}t+i\omega_Lt}\right|^2 d\theta}\). Fig.\ref{fig:axial_solitons} illustrates the fully coherent axial dynamics of the 2D soliton, revealing the synchronization of modes with different axial numbers.  In addition, we note that the axial repetition rate of the 2D soliton \(\Delta\nu_{\text{ax}}=6\) GHz, which corresponds to the axial free spectral range.

\begin{figure*}[hptp!]
    \centering
    \includegraphics[width=\linewidth]{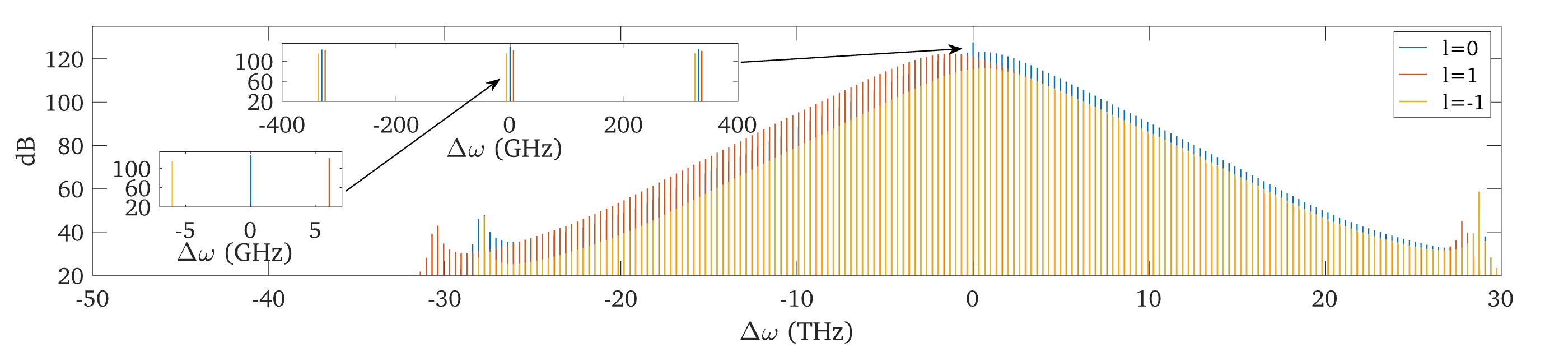}
    \caption{Dual-rate optical frequency comb of the 2D DKS with azimuthal repetition rate \(\Delta\nu_{\text{az}}=329\) GHz and axial sideband repetition rate \(\Delta\nu_{\text{ax}}=6\) GHz. The lines corresponding to different axial numbers $l$ are highlighted in color.}
    \label{fig:comb}
\end{figure*}

At the system output, we obtain an optical frequency comb with a repetition rate of approximately \(\Delta\nu_{\text{az}}=329\) GHz, produced by azimuthal rotation of the 2D soliton. Each comb line has a pair of equally spaced sidebands separated by \(\Delta\nu_{\text{ax}}=6\) GHz, generated by additional slow modulation caused by axial soliton motion, see Fig.\ref{fig:comb} showing the resulting dual-rate spectrum.


{\it Conclusions.}
Usually, microresonator platforms support the generation of one-dimensional DKS within a single-mode family. Simultaneous generation of multiple 1D DKS using multiple or modulated pump of different spatial modes is also possible, enabling multiplexing of two or more frequency combs \cite{lucas2018spatial}. In addition, microresonators support the generation of a limited family of vector DKS, which exist in two polarization modes \cite{bao2019orthogonally}. While various physical mechanisms support the formation of 2D optical dissipative solitons \cite{rosanov2026dissipative}, only a limited number of resonator platforms can support their generation and the formation of frequency combs, as recent theoretical studies show \cite{tusnin2023nonlinear}.

In this Letter, we show that, in a cylindrical microresonator with a radius variation, two-dimensional solitons can form in addition to conventional azimuthal solitons with a stationary axial distribution \cite{Kartashov:18,Eadie:25} and axial solitons with a stationary distribution in the cylinder cross-section \cite{Kolesnikova:25}. These coherent structures localized in both azimuthal and axial directions lead to the generation of optical frequency combs with two non-multiple repetition rates, in contrast to \cite{Kolesnikova:22,Dvoyrin:16}, where an axial--azimuthal comb with a repetition rate determined by the axial free spectral range was obtained (see Fig.~\ref{fig:snap}). Three axial modes are sufficient to trigger modulation instability and the formation of coherent structures, similar to three-mode dynamics in a single LLE model \cite{haelterman1992low}. The number of axial modes can be increased by proper SNAP geometry design, thereby adding more sidebands to the frequency-comb spectra of the 2D DKS. Given recent experimental observations of azimuthal Kerr combs \cite{Eadie:25}, optomechanical oscillations \cite{deep2025cavity}, and topological effects \cite{fried2026multiband} in cylindrical microresonators, we believe that the SNAP platform will enable experimental generation of the predicted 2D DKS states.

Our key finding is a synchronization mechanism and energy transfer between families of azimuthal modes with different axial numbers, leading to the formation of a 2D DKS localized on the microresonator surface, generated by a single-mode pump. Although the free spectral ranges associated with the azimuthal modes initially differ due to their dependence on the axial mode number, the system spontaneously synchronizes the soliton group velocities of the azimuthal coherent states, that resembles recently observed synchronization of solitons in coupled \(\text{Si}
_3\text{N}_4\) Vernier microresonators \cite{yuan2023soliton}. This leads to a common azimuthal repetition rate. Consequently, at the system output, the optical frequency comb exhibits two distinct repetition rates: one set by the azimuthal modes and the other by the axial modes. 

Importantly, the structure of both mode families depends on the cylinder dimensions. Fine tuning of the cylinder radius, for example via compression using a piezoelectric actuator \cite{Papp2013}, leads to simultaneous tuning of both the axial and azimuthal FSRs. For instance, a radius change on the order of 10 nm results in a shift of the azimuthal FSR by approximately 30 MHz. In contrast, the axial FSR, according to Eq.~\eqref{Z_eq} and accounting for the Poisson effect, changes by more than 1 GHz. Since detection of the axial FSR in the gigahertz range is straightforward, simultaneous control of both FSRs enables the control and stabilization of high-frequency microwave comb via stabilization of the lower-frequency comb. This, in turn, opens a new approach to frequency comb generation and may offer advantages for precision spectroscopy and metrology applications.

{\it Acknowledgments.}
The study was supported by the Ministry of Education and Science of the Russian Federation (FSUS-2021-0015).

\bibliography{apssamp}

\end{document}